\documentclass[conference]{IEEEtran}

\usepackage{graphicx}
\usepackage{epsf}

\usepackage{epsfig}

\usepackage{epstopdf}

\newtheorem{definition}{\bf Definition}

\usepackage{ifpdf}

\usepackage{cite}

\ifCLASSINFOpdf
\else
\fi

\usepackage[cmex10]{amsmath}
\usepackage{algorithmic}
\usepackage{array}
\usepackage{mdwmath}
\usepackage{mdwtab}
\usepackage{eqparbox}
\usepackage[tight,footnotesize]{subfigure}
\usepackage{fixltx2e}
\usepackage{stfloats}
\usepackage{url}
\usepackage{caption}
\usepackage{amsfonts}

\begin{document}

\title{Spectrum Sharing in Cooperative Cognitive Radio Networks: A Matching Game Framework}
\author{\IEEEauthorblockN{Nima Namvar, Fatemeh Afghah}
\IEEEauthorblockA{Department of Electrical and Computer Engineering\\
North Carolina A\&T state University, Greensboro, NC 27411\\
Email: nnamvar@aggies.ncat.edu, fafghah@ncat.edu}}

\maketitle

\begin{abstract}
Dynamic spectrum access allows the unlicensed wireless users (secondary users) to dynamically access the licensed bands from legacy spectrum holders (primary users) either on an opportunistic or a cooperative basis. In this paper, we focus on cooperative spectrum sharing in a wireless network consisting of multiple primary and multiple secondary users. In particular, we study the partner-selection and resource-allocation problems within a matching theory framework, in which the primary and secondary users aim at optimizing their utilities in terms of transmission rate and power consumption. We propose a distributed algorithm to find the solution of the developed matching game that results in a stable matching between the sets of the primary and secondary users. Both analytical and numerical results show that the proposed matching model is a promising approach under which the utility functions of both primary and secondary users are maximized.

\end{abstract}

%
\IEEEpeerreviewmaketitle

\section{Introduction} 
Owning to the ever-growing spectrum demand by the recent wireless technologies as well as the inefficiency of traditional  static spectrum assignment policy, spectrum scarcity has become a critical challenge in wireless communication networks \cite{FCC}. Studies show that the traditional static spectrum assignment policy is highly inefficient observing the fact that the allocated bandwidths remain unused by the licensed users, for a considerable amount of time. Indeed, the spectrum utilization can vary in the range of $15\%$ to $85\%$ depending on the geographical distribution and time \cite{85}. Dynamic spectrum sharing is seen as a promising approach toward alleviating spectrum scarcity in wireless networks \cite{DSS}. It allows the unlicensed or secondary users (SUs) to dynamically access the licensed bands of the legitimate primary users (PUs) in exchange for functional or pecuniary compensation.

Generally, there are two scenarios of dynamic spectrum sharing in literature based on the primary users' awareness of the secondary network's presence, namely \emph{common model} and \emph{property-rights model} \cite{Peha}. In the common model, the primary users are oblivious to the secondary network's presence and behave as if there is no secondary activities. The secondary users sense the radio environment in search for spectrum holes and then exploit it in an opportunistic manner. In the property-rights model, it is assumed that the primary users are aware of the secondary network's presence and are willing to lease a portion of their spectrum in exchange for monetary or functional compensations \cite{ZhuHan}, \cite{FatemehJournal} and \cite{SimoneICC}. The functional compensation scenario is called \emph{cooperative} spectrum sharing, which is the focus of this paper.

While the cooperative spectrum sharing carries considerable potential advantages in terms of spectral efficiency, its deployment in wireless networks involves several new technical challenges including interference management, designing incentive-based protocols to encourage the cooperation \cite{FatemehCDC}, \cite{FatemehNDRCS}, selecting the optimal cooperative partner for the nodes, as well as distributed self-organization among the others \cite{ZhuHanBook}.

Recently, game theoretical approaches have been used to study cooperative dynamic spectrum allocation from different aspects. For example, in \cite{stackelberg} the interactions between a PU and a group of SUs is modeled within a Stackelberg game, in which the SUs are granted to exploit the primary link in exchange of monetary as well as functional compensation. Authors in \cite{Simone} propose a Stackelberg game to minimizing the interference among the SUs which cooperate with a single PU. However, most of the existing literature consider a simplified scenario of interaction between a single PU and multiple SUs. More importantly, despite of being of critical importance in cooperative communications, the partner-selection problem has been seldom investigated in the literature.

In this paper, inspired by matching theory, which is a suitable approach to model the interactions among numerous agents with conflicting interests \cite{matching}, we propose a \emph{matching game} framework, in which multiple PUs and multiple SUs interact with one another to select their best possible partner for cooperation in order to optimize their own benefits. The secondary terminals are granted the use of a specific portion of spectrum by the PUs in exchange for cooperative service. The optimum time-frame allocation solution for the cooperative phase and the leased fraction to the SUs is obtained using a Stackelberg game, while the matching scenario provides an optimal answer to the problem of cooperative partner selection. Numerical results confirm that the proposed twofold optimization approach yields considerable gains in terms of transmission rate for both PUs and SUs.

The main contributions of this paper are as follows. i) we optimize the time sharing model for \underline{multiple} PUs and multiple SUs using a Stackelberg game; ii) we propose a novel matching game that captures the preferences of both primary and secondary users to select their optimum partner in a general network consisting of multiple PUs and multiple SUs; iii) we propose a distributed algorithm to solve the matching game that yields a stable matching between the sets of primary and secondary users.

The rest of this paper is organized as follows: In section II, the system model for the proposed scenario is presented. In section III, the problem of spectrum sharing is modeled in the framework of matching theory and a novel algorithm for spectrum sharing is proposed. The simulation results are provided in Section IV. 

\section{System Model} 
Consider a cognitive radio network with $N$ primary and $M$ secondary users. Let $\mathcal{N}=\{PT_i,PR_i\}_{i=1}^N$ and $\mathcal{M}=\{ST_i,SR_i\}_{i=1}^M$ show the set of $N$ primary transmitter-receiver pairs and the set of $M$ secondary transmitter-receiver pairs, respectively. The primary users (PUs) are the spectrum's owners and have the right of interference-free communications in their allocated frequency bands. The secondary users (SUs), on the other hand, seek to obtain the transmission opportunities through negotiation with the PUs. Each primary transmitter can employ one secondary transmitter as a cooperative relay and in turn provide it with the chance of spectrum access. This coordination will improve the quality of service (QoS) for the PUs by exploiting spatial diversity particularly in the cases of poor channel conditions in the primary user's links.

As shown in Figure 1, each time slot is divided into three subslots according to two variables $\alpha$ and $\beta$.  Each $PT_i$ broadcasts its data in the first subslot of duration $1-\alpha$ unit time, and in the second subslot of duration $\alpha\beta$ the selected SU $j$ forwards the message to the corresponding $PR_i$. The third subslot of duration $\alpha(1-\beta)$ unit time is allocated to the secondary transmitter-receiver pair $(ST_j, SR_j)$ for communicating their own data. It is also assumed that all primary transmitters have the same transmission power, $P_{P}$. However, the secondary users are able to adjust their transmission power $P_{S}$ in the range of $[0, P_{max}]$ according to their target utility. We note that the secondary users are constrained to use the same power during the cooperation phase and transmitting their own traffic (phases II and III in Figure 1). This regulation is enforced to assure that the secondary transmitters are trustable in the sense that they treat the primary's data the same as their own \cite{Simone}.

The wireless channels between the nodes are modeled as independent and identically distributed (iid) Rayleigh random variables. For tractability, it is assumed that the instantaneous channel state information (CSI) is available at the primary side. The secondary users are only aware of the channel conditions within the secondary network. All communication channels suffer from block fading which is constant during a time-slot, but can vary over different slots. The notations for channel gains between different nodes are denoted in table I.

\begin{figure}
  \begin{center}
    \includegraphics[width=8cm]{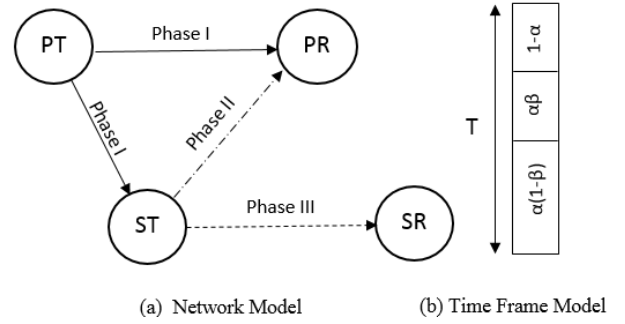}
   \caption{System Model}\vspace{-1cm}
  \end{center}
\end{figure}

\subsection{Transmission Rate for the PUs and SUs} 
Each PU may decide to employ a secondary transmitter as a relay if cooperation improves its transmission rate with respect to the direct transmission. Let $R^{C}_P$ and $R^{NC}_P$ denote the transmission rate of primary link in cooperative and non-cooperative modes, respectively. $R^{NC}_P$ is given by: \vspace{-0.3cm}

\begin{equation}\label{non-cooperative}
  R^{NC}_{P_i}(P_{P_i})=\log(1+\frac{|h_{P_i}|^2P_{P_i}}{N_0}), \hspace{1cm} \forall i\in \mathcal{N}
\end{equation} \vspace{-0.5cm}

Assuming decode-and-forward relaying, the transmission rate of primary link in cooperative mode can be calculated as follows. In the first subslot of duration $1-\alpha_{ij}$, the $PT_i$ transmits to $ST_j$ with the rate $r_1$, which is equal to $\log(1+\frac{|h_{P_iS_j}|^2P_{P_i}}{N_0})$. In the second subslot of duration $\alpha_{ij} \beta_{ij}$, the $PT_i$ is silent and the $ST_j$ transmits to $PR_i$ with the rate $r_2$, which is equal to $\log(1+\frac{|h_{S_jP_i}|^2P_{S_j}}{N_0})$. The overall transmission rate is equal to the minimum data rate of these two phases because the transmission rate is dominated by the worst channel. Therefore, for any $i\in \mathcal{N}$ and $j\in \mathcal{M}$, $R^{C}_{P_i}$ is given by: \vspace{-0.4cm}

\begin{equation}\label{primary rate}
  R^{C}_{P_i}(\alpha_{ij},\beta_{ij},P_{S_j})= \min \text{\{}(1-\alpha_{ij}) r_1, \alpha_{ij}\beta_{ij} r_2\text{\}}.
\end{equation}

Finally, the transmission rate of the secondary link $j$ when it cooperates with PU $i$ can be directly calculated as the following equation: \vspace{-.6cm}

\begin{equation}\label{Secondary rate}
  R_{S_j}(\alpha_{ij},\beta_{ij},P_{S_j})= \alpha_{ij}(1-\beta_{ij})\log \left(1+\frac{|h_{S_j}|^2P_{S_j}}{N_0}\right).
\end{equation}

We note that the cooperation happens if and only if $R^{C}_P>R^{NC}_P$. However for each PU $i$ cooperating with SU $j$, $R^{C}_{P_i}$ depends on $\alpha_{ij}$, $\beta_{ij}$, and $P_{S_j}$ which may vary for any SU $j$. In the next section we derive the optimum values of these variables for all user pairs $(i,j)\in \mathcal{N}\times\mathcal{M}$ which maximize the transmission rates of both users. Having determined these three essential variables, the PUs and the SUs can conclude whether cooperation is beneficial for them. Given the possible cooperating agents for each user, we analyze the problem of partner-selection within a matching game framework.

\begin{table}\label{notations}
\caption{Notations for Channel Gains} \vspace{-.1cm}
\centering
\begin{tabular}{|c|c|}
\hline
Notation & Physical Meaning \\
\hline
$h_{P_i}$             & primary link's channel gain between $PT_i$ and $PR_i$    \\
$h_{S_j}$             & secondary link's channel gain between $ST_j$ and $SR_j$   \\
$h_{P_iS_j}$          & channel gain between $PT_i$ and $ST_j$   \\
$h_{S_jP_i}$          & channel gain between $ST_j$ and $PR_i$   \\
$N_0$             & noise power   \\
\hline
\end{tabular}
\end{table}

\subsection{Optimal values of $\alpha$ and $\beta$}
In this section, we derive the optimal values of $(\alpha_{ij},\beta_{ij})$ and $P_{S_j}$ for all $(i,j)\in \mathcal{N}\times\mathcal{M}$ using a Stackelberg game framework in which the PUs and SUs play the role of the leaders and followers, respectively. Each PU $i$ seeks to maximize its cooperative rate defined in (\ref{primary rate}) by selecting optimum values for $\alpha_{ij}$ and $\beta_{ij}$ (the PU's strategies) with the knowledge of the effect of its decision on the SU $j$'s strategy $(P_{S_j})$. We define the utility of primary users as their achievable transmission rate:

\begin{equation}\label{utility-primary}
 U_{P_i}(\alpha_{ij},\beta_{ij},P_{S_j})=\max(R^C_{P_i},R^{NC}_{P_i}).
\end{equation}

On the other hand, the SUs attempt at maximizing their achievable rate (\ref{Secondary rate}) under a reasonable cost of energy. We define the following utility function of SU $j\in \mathcal{M}$ as its achievable transmission rate minus its cost of energy \cite{Simone}. \vspace{-.5cm}

\begin{equation}\label{utility-secondary}
  U_{S_j}(\alpha_{ij},\beta_{ij},P_{S_j})=R_{S_j}(\alpha_{ij},\beta_{ij},P_{S_j})-\alpha_{ij} CP_{S_j};
\end{equation}
in which $C$ is the cost per unit transmission energy. Given $\alpha_{ij}$ and $\beta_{ij}$, the secondary user $j$ has a unique best strategy $P^*_{S_j}$, which can be found as:
\begin{equation}\label{secondary best power}
  P^*_{S_j}=arg  \smash{\displaystyle\max_{P_{S_j}\in [0,P_{max}]}} U_{S_j}(\alpha_{ij},\beta_{ij},P_{S_j})
\end{equation}

We note that $\alpha_{ij}$ appears in (\ref{utility-secondary}) as a multiplying coefficient and therefore, it does not affect the strategy of secondary user in (\ref{secondary best power}). Consequently, $P^*_{S_j}$ is just a function of $\beta_{ij}$ and can be shown by $P^*_{S_j}(\beta_{ij})$. As the leader of the game, PUs maximize their rates defined in (\ref{primary rate}) by selecting the appropriate values for $\alpha_{ij}$ and $\beta_{ij}$ with the knowledge of their decisions on the strategy of all potential secondary relays ($P_{S_j}$). We note that parameter $\beta_{ij}$ is present only in the second term of (\ref{primary rate}) and therefore, it can be optimized independently as:

\begin{equation}\label{beta}
  \beta^*_{ij}=arg \smash{\displaystyle\max_{0\leq\beta_{ij}\leq1}} \beta\log(1+\frac{|h_{S_jP_i}|^2P^*_{S_j}(\beta_{ij})}{N_0})
\end{equation}

Given $\beta^*_{ij}$, the optimum value $\alpha^*_{ij}$ can be easily computed noting that the cooperative primary rate in (\ref{primary rate}) is the minimum of a decreasing function of $\alpha_{ij}$, $(1-\alpha_{ij}) r_1$, and an increasing function of it, $\alpha_{ij}\beta_{ij} r_2$ which is maximized when the two functions are equal. Solving the equation which results from this condition gives rise to:

\begin{equation}\label{alpha}
  \alpha^*_{ij}=\frac{r_1}{r_1+\beta^*_{ij} r_2}
\end{equation}

Therefore, by substituting (\ref{alpha}) into (\ref{primary rate}), the primary rate in cooperative mode reads: \vspace{-.5cm}

\begin{equation}\label{optimal rate}
  R^{C*}_{P_i}(\alpha^*_{ij},\beta^*_{ij}, P_{S_j})=\frac{\beta^*_{ij} r_1 r_2(P_{S_j})}{r_1+\beta^*_{ij} r_2(P_{S_j})}
\end{equation}

\section{SPECTRUM SHARING AS A MATCHING GAME} 

In the previous section, we derived the optimum values of the time sharing model parameters, $\alpha^*_{ij}$ and $\beta^*_{ij}$, for all pairs $(i,j)\in \mathcal{M}\times\mathcal{N}$. In this section, we study the problem of partner-selection in the cooperative scenario under consideration. Indeed, given the optimum time-sharing model for all possible PU-SU pairs in the network, we are going to find out that when the cooperation is profitable for PUs and what the optimum approach is to assign the SUs to the PUs so as to optimize their utilities.

Originally stems from economics, matching theory \cite{matching} is a suitable mathematical framework to analyze and optimize the problem of partner-selection among two groups of players with conflicting interest. Merits of the stable matching framework lie in the competitiveness of outcomes, generality of the preferences, efficiency and simplicity of its algorithmic implementations, and most importantly, its overall practicality \cite{marriage}. In particular, its advantage over other analytical and numerical optimization methods becomes more evident when the number of decision parameters or the number of players increases beyond a limit where the optimization approaches prove to be unfeasible due to tremendous computational complexity \cite{Nima}.

In this section, we formulate the cooperative spectrum sharing problem as a one-to-one matching game between the set of PUs and the set of SUs to solve the partner-selection problem in the proposed scenario. We analyze the existence of a stable matching and also study its optimality. Let's consider two disjoint sets of $\mathcal{N}$ and $\mathcal{M}$, the primary and secondary users, respectively. Each user has a complete and transitive preference over the users on the other side. We use $\succ_i$ to denote the ordering relationship of agent $i$. For example, $j\succ_i j'$ means that $i$ prefer $j$ over $j'$.

\begin{definition}
A matching $\mu$ is a function from $\mathcal{M}\times\mathcal{N}$ to itself such that $\forall n\in \mathcal{N}$ and $\forall m\in \mathcal{M}$: I. $\mu(n)=m$ if and only if $\mu(m)=n$; and II. $\mu(n)\in \mathcal{M}\cup\emptyset$ and $\mu(m)\in \mathcal{N}\cup\emptyset$.
\end{definition}
This implies that the outcome matches the agents on one side to those on the other side, or to the empty set. The agents’ preferences over outcomes are determined solely by their preferences for their own partners in the matching. To solve a matching game, one suitable concept is that of a stable matching.
\begin{definition}
A matching $\mu$  is blocked by the PU-SU pair (i,j) if $\mu(i)\neq j$ and if $i\succ_j \mu(j)$ and $j\succ_i \mu(i)$. A one-to-one matching is \emph{stable} if it is not blocked by any PU-SU pair.
\end{definition}

We capture the preferences of the primary and secondary users by introducing well-defined utility functions. Based on these utility functions, we analyze the existence of stable matching between the primary and secondary users as the desired outcome of the spectrum sharing problem. The secondary users aim at maximizing their own transmission rates under a reasonable cost of energy consumption according to their utility defined in (\ref{utility-secondary}). On the other hand, the motivation of the PUs to participate in cooperation is to improve their quality of experience (QoE) using spatial diversity. Therefore, for the PUs, we assume that the utility is the transmission rate which is achieved by cooperation, i.e. $R^C_P$ defined in (\ref{primary rate}). Given the utility functions of the SUs and PUs, in the next section we propose an efficient algorithm for solving the game that can find a stable matching between primary and secondary users.

\subsection{Proposed Algorithm}
To solve the formulated game, we propose a novel distributed algorithm shown in Table II. Suppose that all the SUs are initially not associated to any PU. The SUs send their profile information including their CSI and $P_{max}$ to the available PUs. Each PU $i$, on the other side, feeds back the SUs by the ordered pair $(\alpha^*_{ij},\beta^*_{ij})$. Furthermore, each SU selects its strategy (transmit power) according to the time allocation parameters for all the PUs. Given the strategies of the PUs, each SU sends a request of cooperation to its most preferred PU. Among the SU applicants, the PU only keeps the list of SUs who are capable of offering a transmission rate higher than that of the direct path. Formally, for any $i\in \mathcal{N}$ and $j\in \mathcal{M}$, we define the following discriminator function:

 \begin{equation}\label{rate condition}
 DF_i(j)=
\begin{cases}
1,& \hspace{-.1cm} \text{if} \hspace{.3cm} R^C_{P_i}(\alpha^*_{ij},\beta^*_{ij}, P_{S_j})>R^{NC}_{P_i}
\\0, & \hspace{-.1cm} \text{Otherwise }\\
\end{cases}
\end{equation}
Each PU $i$ computes $DF_i(j)$ for all the secondary candidates and only accepts those ones which yield $DF_i(j)=1$. The rest of the SUs will be rejected by PU $i$. Then, each $PU$ ranks all the acceptable SU applicants based on its utility defined in (\ref{primary rate}).
Upon ranking the acceptable SUs, the PU feeds back the awaiting SUs with its decision about the admitted cooperator. The SUs who have been rejected in the former phase will apply to their next favorite PU and the PUs compare the new applicants with their current temporary partner (if any) and again select the most preferred one among them. This procedure continues until all the SUs assigned to one of the PUs or further proposals are impossible.

Here, we note that the outcome of one-to-one matching markets is optimum for the set of players who make the proposals \cite{matching} which in our model, is the set of SUs. The proposed algorithm yields an stable matching between the two sets of $\mathcal{M}$ and $\mathcal{N}$ for any initial preference functions and the resultant matching is optimal from the SUs' point of view. The deferred acceptance method which is used in stage II of the algorithm converges for any initial conditions. The mathematical proof of stability amounts and we refer the reader to \cite{matching} and \cite{GaleShapley}. However, the stability of the outcome matching is intuitive since by introducing the null set $\emptyset$ as a possible partner, each user has the option to stay unmatched rather than being matched to partner which does not satisfy its utility. Therefore, no blocking pair will emerge in the iterative deferred acceptance algorithm in stage II because each SU only propose to those PUs who satisfy its utility. As a result, the matching includes no blocking pairs and therefore, it is stable.

 \begin{table}[t] 
  \scriptsize
  \centering
  \caption{Proposed Algorithm For The Matching Game} \vspace*{-1em}
    \begin{tabular}{p{8.5cm}}
      \hline \vspace*{.5em}
\textbf{Input:} Utilities and the preferences of each set $\mathcal{M}$ and $\mathcal{N}$ \\
\textbf{Output:} Stable matching between the primary and secondary users \\
\\
\textbf{Initializing}: All the SUs are matched with the null set $\emptyset$ \\
\textbf{Stage I}: \textbf{Preference Lists Composition}
\begin{itemize}
  \item PUs and SUs exchange their profile information
  \begin{itemize}
  \item Each PU $i$ computes $\alpha^*_{ij}$ and $\beta^*_{ij}$ according to its requirements
  \item Each SU $j$ selects its transmission power $P_{S_j}$ corresponding to $\alpha^*_{ij}$ and $\beta^*_{ij}$ for each $i\in \mathcal{N}$
  \end{itemize}
  \item Each PU $i$ sorts the set of acceptable candidate SUs with $DF_i(j)=1$
  \item SUs sort the PUs based on their preference functions (\ref{utility-secondary})
\end{itemize} \\
\textbf{Stage II}: \textbf{Matching Evaluation} \\
\hspace*{1em}\textbf{while:} $\mu^{(n+1)}\neq \mu^{(n)}$
\begin{itemize}
  \item Each SU $j$ applies to its most preferred PU
  \item Each PU $i$ accepts the most preferred applicant and rejects the rest
\end{itemize}
\hspace*{2em} \textbf{Repeat}\\
\hspace*{4em} $\bullet$ Each rejected SU applies to its next preferred PU \\
\hspace*{4em} $\bullet$ Each PU updates its partner considering the new applicants and the \\
\hspace*{5em} pervious partner\\
\hspace*{2em} \textbf{Until:} Each SU are either assigned to one of the PUs or rejected by all of the \\
\hspace*{5em} PUs  \\
\hspace*{1em}\textbf{end} \\
   \hline
    \end{tabular}\label{tab:algo}

\end{table}

\section{Simulation Results}

\begin{figure}
  \begin{center}
    \includegraphics[width=9cm]{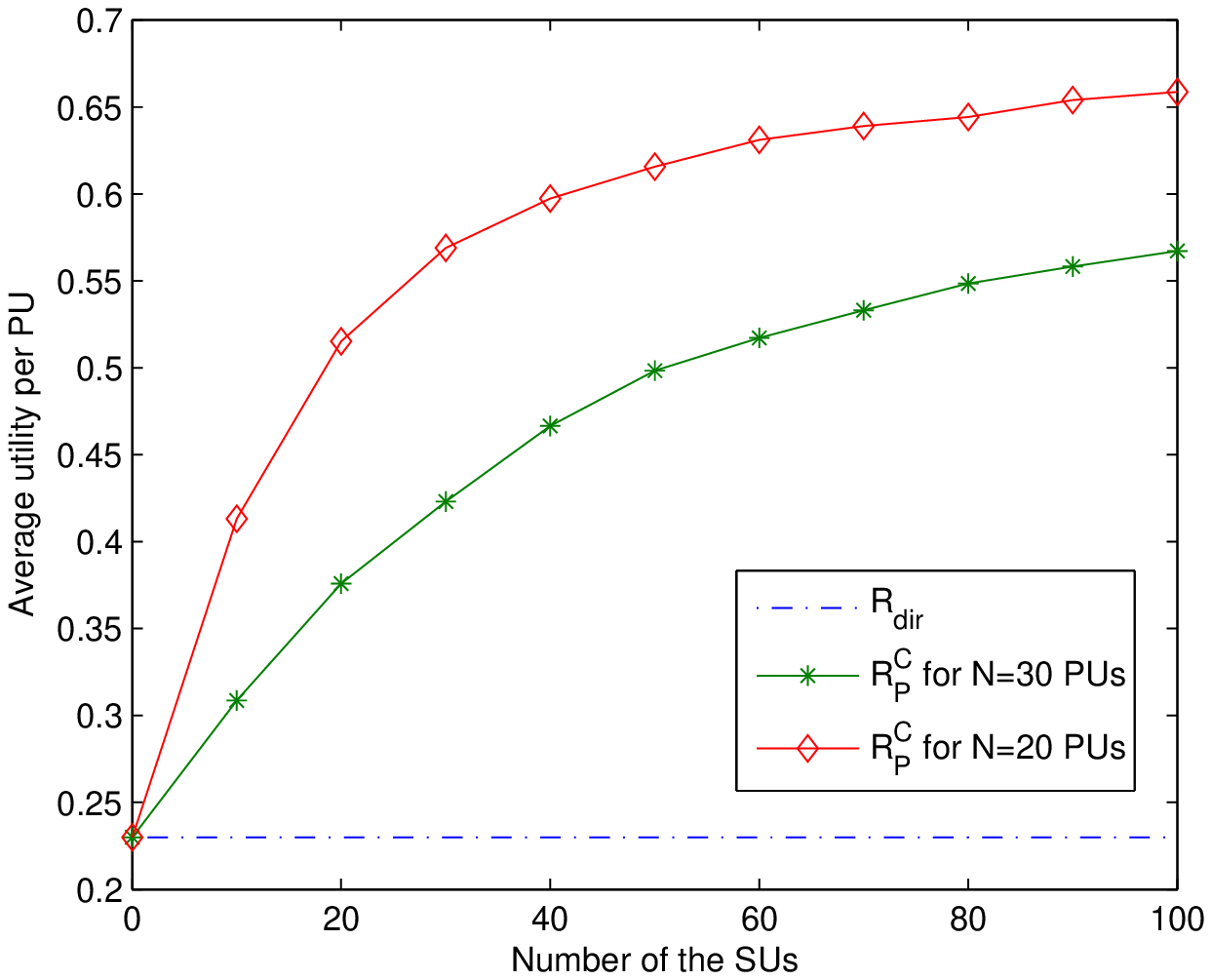}
   \caption{Average utility per PU for different number of SUs with a network size of $N=20$ and $N=30$ PUs}
   \label{sim1}
  \end{center}
\end{figure}

\begin{figure}
  \begin{center}
    \includegraphics[width=9cm]{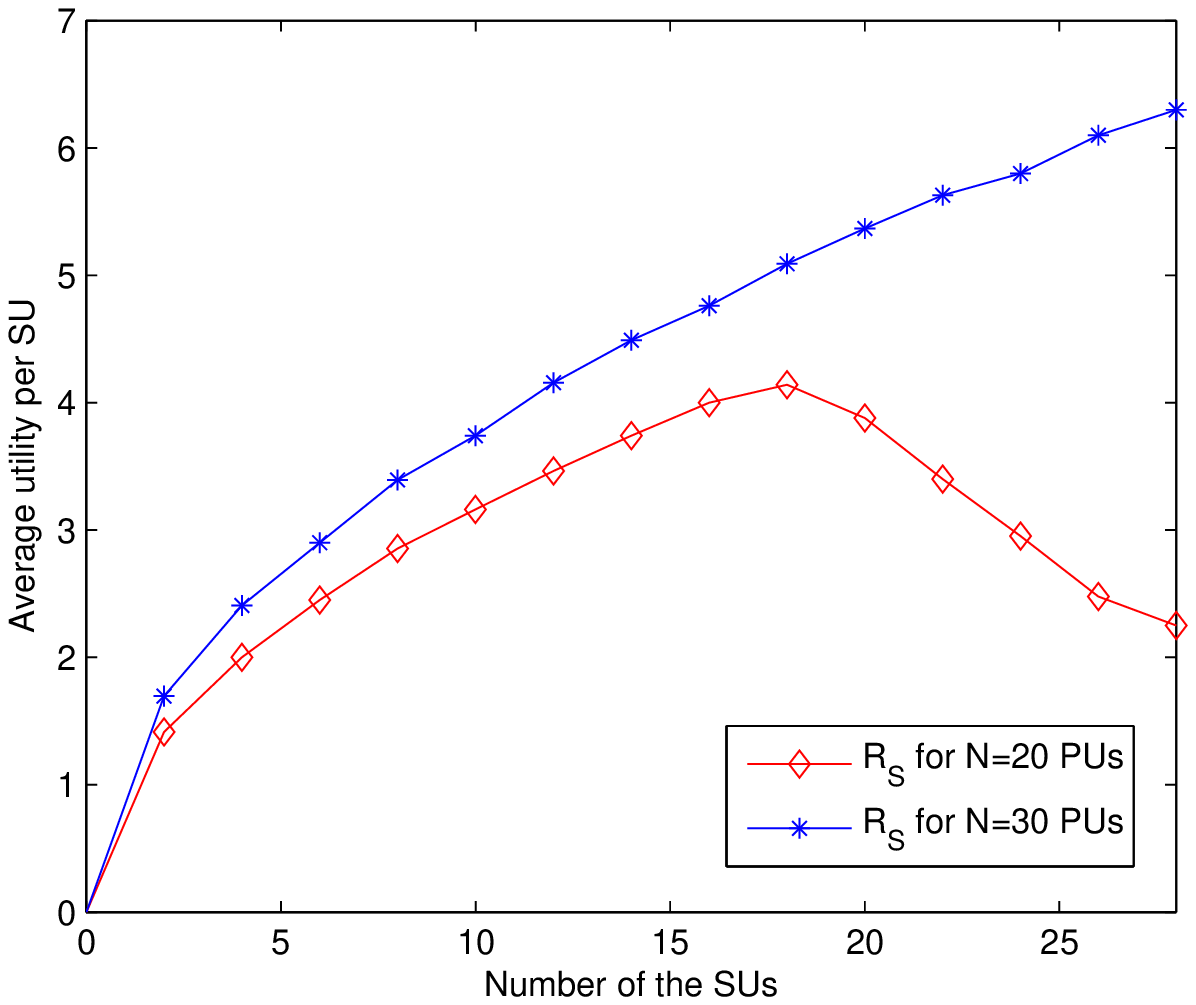}
   \caption{Average utility per PU for different number of SUs with a network size of $N=20$ and $N=30$ PUs}
   \label{sim2}
  \end{center}
\end{figure}

For our simulations, we consider a network consisting of $N$ primary users and $M$ secondary users. The wireless fading channels are i.i.d. and have Rayleigh distribution with the scale parameter $\sigma=0.5$.

Figure \ref{sim1} shows the average achievable rate per primary user as a function of the number of SUs for two network size of $N=30$ and $N=20$ PUs. As the number of the SUs increases, more PUs get the chance to access a cooperating relay and the average achievable rate will increase; specially when the primary network's size is smaller. We note that $R^{NC}_{P}$ is independent of the secondary network and therefore, it is constant. Figure \ref{sim1} shows that the proposed cooperative matching approach yields considerable gain over non-cooperative scenario.

Figure \ref{sim2} shows the average utility per secondary user as a function of the number of SUs for two network size of $N=30$ and $N=20$ PUs. It shows that as the number of SUs increases, the average utility per SU will also increase because more SUs get the chance to access a primary link for communication. However, as the number of SUs increases, finding a primary link becomes more competitive and beyond some point, the average utility of SUs starts to decrease. Figure \ref{sim2} shows that for a network size of $N=20$ PUs, the average utility for SUs starts to decrease after introducing more than $M=20$ SUs in the network. Indeed, increasing the number of SUs beyond $M=20$ will inevitably result in having some unmatched SUs in the network with utility $0$ which consequently, decreases the average utility per SU. We can conclude that increasing the number of SUs beyond the network size $N$ makes the average utility per SU to approach zero gradually.

\section{Conclusions}
In this paper, we have proposed a novel cooperative spectrum sharing approach for a wireless network consisting of multiple primary and secondary users. By introducing well-designed utility functions, we modeled the problem of partner-selection as a one-to-one matching game which optimizes the utility of secondary network. To solve the presented matching game, we have proposed a distributed algorithm that converges to a stable matching between the set of primary users and the set of secondary users. Simulation results show that the proposed cooperative approach yields considerable gains in terms of transmission rate compared to that of the non-cooperative scenario for the primary users.

\bibliography{myref}

\begin{thebibliography}{10}
\providecommand{\url}[1]{#1}
\csname url@samestyle\endcsname
\providecommand{\newblock}{\relax}
\providecommand{\bibinfo}[2]{#2}
\providecommand{\BIBentrySTDinterwordspacing}{\spaceskip=0pt\relax}
\providecommand{\BIBentryALTinterwordstretchfactor}{4}
\providecommand{\BIBentryALTinterwordspacing}{\spaceskip=\fontdimen2\font plus
\BIBentryALTinterwordstretchfactor\fontdimen3\font minus
  \fontdimen4\font\relax}
\providecommand{\BIBforeignlanguage}[2]{{%
\expandafter\ifx\csname l@#1\endcsname\relax
\typeout{** WARNING: IEEEtran.bst: No hyphenation pattern has been}%
\typeout{** loaded for the language `#1'. Using the pattern for}%
\typeout{** the default language instead.}%
\else
\language=\csname l@#1\endcsname
\fi
#2}}
\providecommand{\BIBdecl}{\relax}
\BIBdecl

\bibitem{FCC}
FCC, ``Report of the spectrum efficiency working group,'' \emph{FCC Spectrum
  Policy Task Force, Techical Report}, Nov 2002.

\bibitem{85}
I.~F. Akyildiz, W.~Lee, M.~C. Vuran, and S.~Mohanty, ``Next generation/dynamic
  spectrum access/cognitive radio wireles networks: A survey,'' \emph{Computer
  Networks: The International Journal of Computer and Telecommunications
  Networking}, vol.~50, no.~13, p. 2127–2159, May 2006.

\bibitem{DSS}
Q.~Zhao and B.~Sadler, ``A survey of dynamic spectrum access,'' \emph{Signal
  Processing Magazine, IEEE}, vol.~24, no.~3, pp. 79--89, May 2007.

\bibitem{Peha}
J.~Peha, ``Approaches to spectrum sharing,'' \emph{Communications Magazine,
  IEEE}, vol.~43, no.~2, pp. 10--12, Feb 2005.

\bibitem{ZhuHan}
D.~Niyato, E.~Hossain, and Z.~Han, ``Dynamics of multiple-seller and
  multiple-buyer spectrum trading in cognitive radio networks: A game-theoretic
  modeling approach,'' \emph{Mobile Computing, IEEE Transactions on}, vol.~8,
  no.~8, pp. 1009--1022, Aug 2009.

\bibitem{FatemehJournal}
F.~Afghah and A.~Razi, ``Game theoretic study of cooperative spectrum leasing
  in cognitive radio networks,'' \emph{International Journal of Handheld
  Computing Research (IJHCR)}, vol.~5, no.~2, pp. 61--74, June 2014.

\bibitem{SimoneICC}
O.~Simeone, J.~Gambini, Y.~Bar-Ness, and U.~Spagnolini, ``Cooperation and
  cognitive radio,'' in \emph{Communications, 2007. ICC '07. IEEE International
  Conference on}, June 2007, pp. 6511--6515.

\bibitem{FatemehCDC}
F.~Afghah, M.~Costa, A.~Razi, A.~Abedi, and A.~Ephremides, ``A reputation-based
  stackelberg game approach for spectrum sharing with cognitive cooperation,''
  in \emph{Decision and Control (CDC), 2013 IEEE 52nd Annual Conference on},
  Dec 2013, pp. 3287--3292.

\bibitem{FatemehNDRCS}
F.~Afghah and A.~Razi, ``Cooperative spectrum leasing in cognitive radio
  networks,,'' in \emph{National Wireless Research Collaboration Symposium},
  May 2014, pp. 106--111.

\bibitem{ZhuHanBook}
E.~Hossain, D.~Niyato, and Z.~Han, \emph{Dynamic Spectrum Access and Management
  in Cognitive Radio Networks}.\hskip 1em plus 0.5em minus 0.4em\relax
  Cambridge, U.K.: Cambridge Univ. Press, 2009.

\bibitem{stackelberg}
J.~Zhang and Q.~Zhang, ``Stackelberg game for utility-based cooperative
  cognitiveradio networks,'' in \emph{Proceedings of the 10th ACM international
  symposium on Mobile ad hoc networking and computing}, May 2009.

\bibitem{Simone}
O.~Simeone, I.~Stanojev, S.~Savazzi, Y.~Bar-Ness, U.~Spagnolini, and
  R.~Pickholtz, ``Spectrum leasing to cooperating secondary ad hoc networks,''
  \emph{Selected Areas in Communications, IEEE Journal on}, vol.~26, no.~1, pp.
  203--213, Jan 2008.

\bibitem{matching}
A.~Roth and M.~Sotomayor, ``Two-sided matching: A study in game theoretic
  modeling and analysis,'' \emph{Cambridge Univ. Press}, 1992.

\bibitem{marriage}
H.~Xu and B.~Li, ``Seen as stable marriages,'' in \emph{INFOCOM, 2011
  Proceedings IEEE}, April 2011, pp. 586--590.

\bibitem{Nima}
N.~Namvar, W.~Saad, B.~Maham, and S.~Valentin, ``A context-aware matching game
  for user association in wireless small cell networks,'' in \emph{Acoustics,
  Speech and Signal Processing (ICASSP), 2014 IEEE International Conference
  on}, May 2014, pp. 439--443.

\bibitem{GaleShapley}
D.~Gale and L.~S. Shapley, ``College admissions and the stability of
  marriage,'' \emph{The American Mathematical Monthly}, vol.~69, no.~1, pp.
  9--15, 1962.

\end{thebibliography}
\bibliographystyle{IEEEtran}

\end{document}